\documentclass[final,3p,times]{elsarticle}

\usepackage{amssymb}
\usepackage{amsmath}

\journal{Journal of Subatomic Particles and Cosmology}

\begin{document}

\begin{frontmatter}



\title{QCD phase structure at high baryon density}

\author[FRIB]{Agnieszka Sorensen}
\affiliation[FRIB]{organization={Facility for Rare Isotope Beams, Michigan State University},
             addressline={640 S.\ Shaw Lane},
             city={East Lansing},
             postcode={48824},
             state={MI},
             country={USA}}

\begin{abstract}
Heavy-ion collisions provide unique access to the properties of QCD matter over a broad range of temperatures and baryon densities. 
We review recent progress in constraining the high-density phase structure of QCD, with particular emphasis on the microscopic inputs required for reliable transport modeling. 
We discuss cluster production, strange-particle interactions, collision dynamics at few-GeV energies, and the isospin dependence of the nuclear interaction, together with their impact on the interpretation of flow, yields, fluctuations, and possible critical signatures. 
\end{abstract}




\end{frontmatter}



\section{Introduction}
\label{sec:introduction}

Uncovering the phase diagram of quantum chromodynamics~(QCD) is one of the core objectives of nuclear physics studies. 
While large regions of the QCD phase diagram remain unconstrained, decades of theoretical and experimental studies have established several important features of its structure or, equivalently, of the dependence of the nuclear matter equation of state~(EOS) on baryon density~$n_B$, temperature~$T$, and isospin fraction.

At $T=0$, isospin-symmetric nuclear matter saturates at density~$n_0 \approx 0.160~\rm fm^{-3}$~\cite{Bethe:1971xm}, where bound nuclear matter coexists with the vacuum, corresponding to the zero-temperature limit of the nuclear liquid-gas first-order phase transition.
Theoretical studies and experimental measurements indicate that this first-order phase transition terminates at the nuclear liquid-gas critical point, located approximately at~$n_B \approx 0.4n_0$ and $T \approx 18~\rm MeV$~\cite{Elliott:2013pna,Wellenhofer:2014hya}.

At low densities and up to about 1.5 times~$n_0$, chiral effective field theory~($\chi$EFT) provides robust calculations of the EOS for both symmetric and asymmetric nuclear matter.
More recently, the reach of $\chi$EFT has been extended to finite temperatures, providing predictions for the EOS up to about $T \approx 20 ~\rm MeV$~\cite{Keller:2020qhx,Keller:2022crb}.

At moderate and high~$n_B$, astronomical observations of pulsars constrain highly isospin-asymmetric, neutron-rich matter at $T =0$.
In particular, the existence of very massive neutron stars~\cite{NANOGrav:2019jur,Fonseca:2021wxt} requires sufficiently large pressure at densities of several times~$n_0$~\cite{Bedaque:2014sqa,Alford:2015dpa,Tews:2018kmu,Fujimoto:2019hxv, Marczenko:2022jhl}. 
On the other hand, at sufficiently high~$n_B$ -- or, equivalently, large baryon chemical potential~$\mu_B$ -- strange degrees of freedom are expected to become relevant.
Their appearance generally tends to soften the EOS~\cite{Oertel:2016bki,Schaffner-Bielich:2008zws}, giving rise to the well-known hyperon puzzle: in many models, the associated softening is sufficiently strong that the resulting EOS cannot support the heaviest observed neutron stars~\cite{Chatterjee:2015pua}. 

Over the last decade, groundbreaking gravitational-wave~(GW) measurements have opened a new avenue for constraining the properties of binary neutron-star systems.
In particular, measurements of the tidal deformability during the inspiral provide complementary constraints on the neutron-star EOS~\cite{De:2018uhw,LIGOScientific:2018cki}.
As instrumentation and analyses advance, the post-merger GW signal may eventually provide access to finite-temperature neutron-rich matter; at present, however, such information remains beyond observational reach.

At $n_B = 0$, or equivalently $\mu_B = 0$, lattice QCD calculations indicate that hadronic matter undergoes a crossover transition to the quark-gluon plasma~(QGP) at a pseudocritical temperature of about~$T_{pc} = 156.5 \pm 1.5~ \rm MeV$~\cite{HotQCD:2018pds} (see also Refs.~\cite{Aoki:2006br,Borsanyi:2020fev}).
Evidence gathered over 30 years of experiments at the Super Proton Synchrotron~(SPS), the Relativistic Heavy Ion Collider~(RHIC), and the Large Hadron Collider~(LHC) strongly indicates that a new state of matter is indeed produced in high-energy heavy-ion collisions~\cite{Heinz:2000bk,BRAHMS:2004adc,PHENIX:2004vcz,PHOBOS:2004zne,STAR:2005gfr,Muller:2012zq}. 

Finally, at asymptotically high temperatures, asymptotic freedom implies that strongly-interacting matter approaches a weakly-coupled, deconfined plasma of quarks and gluons~\cite{Gross:1973id,Politzer:1973fx,Ghiglieri:2020dpq}, while at asymptotically high baryon densities QCD predicts phases dominated by quark degrees of freedom~\cite{Collins:1974ky,Alford:1998mk}.

Outside of these comparatively well-established regimes, the QCD phase diagram remains largely unknown. 
In particular, it is not known whether, as $\mu_B$ is increased from zero, the crossover between hadronic matter and the QGP eventually turns into a first-order phase transition and, consequently, whether the QCD phase diagram contains a critical point~(CP)~\cite{Bzdak:2019pkr}.
Large uncertainties also remain in the isospin dependence of the EOS, affecting both our understanding of neutron-rich matter and the information that can be extracted from observations of neutron stars and their mergers~\cite{Sorensen:2023zkk}.
Both questions concern fundamental properties of QCD matter.

\section{Role of heavy-ion collision experiments and selected constraints}

Relativistic heavy-ion collisions provide access to a substantial region of the QCD phase diagram.
As the collision energy increases, the temperatures reached in the collision generally increase, while the maximum baryon density first rises, peaking at center-of-mass energies per nucleon pair~$\sqrt{s_{\rm NN}}$ of several GeV, and then decreases toward the nearly zero net-baryon density systems created in collisions at top energies available at RHIC and the LHC.
In this way, varying the collision energy provides access to QCD matter over a broad range of~$T$ and~$n_B$.
Notably, heavy-ion collision experiments are the only experimental means for probing the QCD EOS away from the saturation density.
Most moderate and high-energy facilities, such as GSI, RHIC, and the LHC, collide beams of stable nuclei, primarily probing nearly isospin-symmetric nuclear matter.
The isospin-dependence of the EOS can be explored more directly in experiments colliding proton- or neutron-rich nuclei, including unstable beams available at the Facility for Rare Isotope Beams~(FRIB), RIKEN, or RAON.

Information about the high-density QCD phase structure can be extracted by comparing experimental observables with dynamical simulations of the collisions.
These simulations incorporate assumptions about the QCD phase diagram through the QCD EOS as well as other properties of the underlying microscopic dynamics.
The large past, present, and future experimental program in heavy-ion physics now provides a broad and systematically varied data set spanning a wide range of collision energies, system sizes, and colliding species (we note, however, that the coverage is not complete, as highlighted in Ref.~\cite{Naim:2026tvp}).
At the same time, increasingly precise measurements, improved control of experimental systematics, growing computational resources, and advances in emulation and Bayesian inference have made it possible to perform substantially more comprehensive comparisons between models and data.
Together, these developments create the opportunity to place quantitative constraints on the high-density QCD phase structure, including both the EOS and possible critical behavior.

A major objective of beam-energy scan programs is the search for signatures of a first-order phase transition and a possible QCD CP at finite baryon density. 
This search is particularly timely given that recently, several QCD CP predictions have been obtained from lattice QCD studies utilizing calculations at imaginary~$\mu_B$ and Lee-Yang edge singularities~\cite{Mukherjee:2019eou,Connelly:2020pno,Basar:2021hdf,Dimopoulos:2021vrk,Schmidt:2022ogw,Zambello:2023ptp,Clarke:2024ugt,Wan:2025wdg} as well as from functional renormalization group calculations~\cite{Fu:2019hdw,Gao:2020qsj,Gao:2020fbl,Fu:2026qnl}, Dyson--Schwinger methods~\cite{Fischer:2014ata,Gunkel:2021oya}, holographic gauge/gravity calculations~\cite{Hippert:2023bel,Ecker:2025vnb}, and other promising approaches~\cite{Bluhm:2024uhj,Shah:2024img}.
Notably, these predictions place the QCD CP in a similar region of the QCD phase diagram, around $\mu_B \sim 500$--$700~\rm MeV$ and $T \sim 80$--$120~\rm MeV$.

In the experimental search for the CP, particular attention has been devoted to event-by-event fluctuations of conserved charges, whose higher-order cumulants are expected to exhibit enhanced sensitivity to critical dynamics~\cite{Stephanov:1998dy,Stephanov:1999zu,Asakawa:2009aj,Stephanov:2011pb}, and the STAR Collaboration has provided measurements of cumulant observables as a function of collision energy~\cite{STAR:2021iop,STAR:2025zdq}. 
Their interpretation, however, remains challenging: finite-size and finite-time effects, nonequilibrium evolution, global conservation laws, acceptance effects, and noncritical dynamical correlations can all modify the measured cumulants. 
Establishing whether the observed behavior reflects critical dynamics therefore requires quantitative dynamical modeling together with reliable noncritical baselines; see Ref.~\cite{Pradeep:2026ggj} for recent developments in this area.
A complementary approach is to exploit the finite size of the collision system directly. 
In particular, a recent finite-size-scaling analysis of second-order net-proton cumulants yielded a prediction for the location of the QCD CP consistent with the theoretical predictions highlighted above~\cite{Sorensen:2024mry}.
At the same time, the study demonstrated that apparent scaling behavior can also emerge in systems without a CP, with the effect traced to the evolution of the observables along the chemical freeze-out line.

Information on the phase structure need not come from fluctuation observables alone. 
Since a phase transition or rapid crossover can generate characteristic structures in the density dependence of the pressure and speed of sound, constraints on the EOS from particle yields and collective flow can provide complementary information on the underlying QCD phase diagram.
At collision energies around $\sqrt{s_{\rm NN}} \approx 10 ~\rm GeV$ and above, the rapid approach toward local equilibrium makes hybrid descriptions combining relativistic hydrodynamics with hadronic transport particularly useful (for more details, see Ref.~\cite{Du:2026rxk}).
At lower collision energies, however, where the largest net-baryon densities are reached, the interpenetration time of the two nuclei becomes comparable to the overall dynamical timescale and the system can remain far from equilibrium throughout a significant fraction of its evolution.
In this regime, the collision is commonly described using microscopic hadronic transport approaches, including \texttt{UrQMD}~\cite{Bleicher:1999xi,Bass:1998ca}, \texttt{SMASH}~\cite{SMASH:2016zqf,Elfner:2025ojd}, \texttt{GiBUU}~\cite{Buss:2011mx}, and \texttt{JAM}~\cite{Nara:1999dz,Isse:2005nk}.
Notably, using a single dynamical framework across the full collision evolution avoids additional uncertainties associated with matching distinct descriptions of the early, dense, and late stages, as required in hybrid models.
At the same time, extracting reliable EOS constraints from microscopic transport requires a quantitative understanding of other ingredients of the dynamics, including the momentum dependence of mean-field interactions, in-medium scattering, and cluster formation.
This requirement also presents an opportunity: improving these ingredients advances our understanding of the microscopic properties of dense QCD matter while simultaneously reducing the theoretical uncertainties entering the extraction of the high-density EOS.

Constraints on the high-density EOS from heavy-ion collisions have a long history.
A landmark analysis by Danielewicz \textit{et al.}~\cite{Danielewicz:2002pu} compared measurements of collective flow over $\sqrt{s_{\rm NN}}=1.95$--$4.72~\rm GeV$~\cite{Gustafsson:1988cr,EOS:1994kku,E877:1997zjw,E895:2000maf} with \texttt{pBUU} transport calculations~\cite{Danielewicz:1991dh,Danielewicz:1999zn}.
The resulting constraint on the pressure of symmetric nuclear matter, extending approximately over $(2$--$4.5)n_0$, favored an EOS lying between Skyrme-like parametrizations with incompressibilities $K_0=210$ and $300~\rm MeV$.
A later analysis of FOPI elliptic-flow measurements at $\sqrt{s_{\rm NN}}=2.07$--$2.52~\rm GeV$~\cite{FOPI:2011aa}, using \texttt{IQMD} calculations~\cite{Aichelin:1991xy,Hartnack:1997ez}, obtained $K_0=190\pm30~\rm MeV$~\cite{LeFevre:2015paj}, likewise favoring a comparatively soft EOS around and moderately above~$n_0$.
More recent studies have moved beyond comparisons among a small number of predefined EOSs.
Bayesian analyses using flexible density-dependent forms of the EOS, including those of Oliinychenko \textit{et al.}~\cite{Oliinychenko:2022uvy} and Omana Kuttan \textit{et al.}~\cite{OmanaKuttan:2022aml}, demonstrate that heavy-ion data can constrain significantly more general behavior of the EOS at supranuclear densities.
At the same time, these analyses tend to favor values of~$K_0$ that are considerably larger than established empirical expectations, highlighting an important limitation of current extractions: lack of momentum dependence of the nuclear interaction.
Indeed, earlier transport studies incorporated an explicitly momentum-dependent nuclear mean field, motivated by both experiment~\cite{Cooper:1987uy,Hama:1990vr} and microscopic calculations~\cite{Botermans:1990qi}, that becomes increasingly repulsive at large nucleon momenta.
By contrast, while recent Bayesian analyses have allowed substantially greater freedom in the density dependence of the EOS, they do not model the momentum dependence.
Because both effects influence collective flow, neglecting momentum dependence can cause its dynamical impact to be absorbed into the inferred density dependence of the EOS, potentially biasing the extracted high-density pressure toward more repulsive behavior.
A robust determination of the EOS therefore requires constraining its density and momentum dependence simultaneously within a common transport framework.
(See Refs.~\cite{Sorensen:2023zkk,Du:2024wjm} for compilations of the above-described constraints.)

This example illustrates a broader opportunity offered by microscopic transport theory.
Several ingredients that currently limit the precision of EOS extraction are themselves open problems in dense nuclear matter.
Improving their treatment will not only reduce model uncertainties, but also address the underlying microscopic mechanisms that govern the evolution of the collision.

\section{Microscopic dynamics of dense nuclear matter}

In this section, we highlight several recent developments, some of them presented at this conference, that illustrate how experimental observables, together with dynamical modeling, can be used to improve our understanding of the complex reaction dynamics at low collision energies.

\subsection{Cluster production}

One of the most pressing problems in modeling low-energy heavy-ion collisions is the mechanism of cluster, or light nuclei, production.
Traditionally, three families of approaches have been used: kinetic (also referred to as dynamical) production, coalescence, and thermal production.
In kinetic approaches, clusters are introduced as explicit degrees of freedom and participate in the dynamical evolution through corresponding production, scatterings, and decays~\cite{Danielewicz:1991dh,Oliinychenko:2018ugs}.
In coalescence approaches, clusters are instead constructed from nucleons that are sufficiently close in coordinate and momentum space, typically toward the end of the collision evolution~\cite{Ono:2018vht}; more sophisticated prescriptions may additionally impose conditions related to the binding energy of the resulting cluster~\cite{LeFevre:2019wuj,Kireyeu:2024hjo}.
Finally, in thermal description (also known as the statistical hadronization model), cluster abundances are determined statistically at chemical freeze-out together with those of other hadronic species~\cite{Andronic:2010qu,Steinheimer:2012tb}.

Each of these methods has advantages and limitations. 
A significant advantage of kinetic production is that light nuclei participate in the dynamics throughout the evolution.
As highlighted in Ref.~\cite{Oliinychenko:2018ugs}, a cluster can be created and destroyed many times before the system reaches its final state.
When a substantial fraction of nucleons becomes bound in clusters, such processes can in principle modify not only the final cluster yields but also the dynamics of the remaining unbound nucleons.
However, the kinetic treatment becomes progressively more difficult as heavier nuclei are included, given that the number of possible production and breakup channels grows rapidly while many of the corresponding reaction rates are poorly constrained experimentally.
For this reason, practical implementations have so far concentrated primarily on the lightest nuclei, typically deuterons, tritons, and $^3{\rm He}$.
Coalescence provides an alternative approach in which clusters are assembled from their constituents at a late stage of the evolution.
This method can successfully reproduce final yields of multiple species of light clusters, however, by construction it does not take into account the effects of cluster production on the dynamical evolution. 
Thermal models take an even more coarse-grained view, describing observed cluster abundances through statistical production at freeze-out without attempting to resolve the microscopic mechanism by which the bound states are formed.

Notably, the choice of cluster-production mechanism can affect observables that are not themselves cluster observables.
Indeed, Ref.~\cite{Mohs:2020awg} shows that at low collision energies, the predicted proton flow substantially changes depending on how cluster formation is treated.
A reliable description of cluster production is therefore important not only for understanding light nuclei themselves, but also for extracting properties of the underlying nuclear interaction from observables involving unbound nucleons.

Recent measurements provide interesting clues about the microscopic cluster formation mechanism.
The ALICE experiment used femtoscopic correlation measurements to investigate deuteron production~\cite{ALICE:2025byl}.
The analysis indicates that about 89\% of deuterons are produced in reactions of a ``free'' nucleon combining with a nucleon originating from a resonance decay.
In this picture, a short-lived hadronic resonance -- \textit{e.g.}, a \(\Delta\)-resonance -- decays, after which the daughter nucleon combines with another nucleon to form the deuteron. 
Because the pion from the resonance decay remains correlated with the resulting deuteron, the production history can be studied with femtoscopy, where the observed correlations point to a specific dynamical sequence of resonance decay followed by deuteron formation\footnote{In the terminology commonly used in hadronic transport calculations, such a mechanism has much in common with kinetic cluster production, since the deuteron is formed through an explicit reaction involving dynamically produced constituents. 
Consequently, the term ``resonance-induced coalescence,'' used by the ALICE Collaboration, should not be confused with the conventional end-of-evolution phase-space coalescence.}.
These measurements provide valuable guidance for the reaction pathways that could be incorporated into microscopic descriptions of deuteron production.

A complementary perspective is provided by recent ALICE measurements of $^3{\rm He}$ elliptic flow~\cite{ALICE:2026hca}, which reaches values exceeding $0.5$ at high transverse momentum~$p_T$.
Such large anisotropies cannot be interpreted as the elliptic flow of an independently emitted thermal species using only the leading harmonic, since $v_2>0.5$ would render the corresponding truncated azimuthal distribution negative (\textit{i.e.}, unphysical) in certain angular regions.
However, the observed magnitude can arise naturally if the $^3{\rm He}$ nucleus inherits the collective motion of its constituent nucleons through coalescence.
This interpretation should be treated with some caution: coalescence is generally expected to become increasingly important toward the high-$p_T$ end of the spectrum (as is well-established in the case of quark coalescence~\cite{Molnar:2003ff}) and the measurement therefore does not by itself identify the dominant production mechanism over the full phase space.
Nevertheless, together with the femtoscopic measurement, it provides evidence that the phase-space distributions and dynamical histories of the constituent nucleons play an important role in light-nucleus formation, beyond what is captured by a purely statistical production picture.

\subsection{Strange particle interactions}

Many unknowns remain regarding the description of interactions of strange hadrons with nuclear matter.
In relativistic mean-field models, the couplings of hyperons to the vector mean field are often guided by simple light-quark counting.
In that case, $\Lambda$ and $\Sigma$~baryons couple to the vector nuclear mean field with a reduced factor of~$2/3$, and the $\Xi$~baryon couples with a factor of~$1/3$.
This motivates implementations in many transport codes in which the corresponding mean-field interactions are scaled by these factors.
However, such simple scaling prescriptions should be treated with caution, since hypernuclear spectroscopy constrains the total $\Lambda$ single-particle potential around~$n_0$ at approximately $U_{\Lambda}(n_0) \approx - 28~\rm{MeV}$~\cite{Millener:1988hp,Yamamoto:1988qz}, substantially less attractive -- by a factor of about~$1/2$ -- than the nucleon single-particle potential.
In relativistic mean-field models, this is typically accommodated by adjusting the scalar hyperon couplings in addition to the vector couplings.

Even less is known about hyperon interactions at densities above saturation.
While microscopic calculations based on chiral effective field theory indicate that three-body $\Lambda NN$ interactions can generate substantial repulsion with increasing density, the strength of this high-density repulsion remains poorly constrained~\cite{Gerstung:2020ktv,Kohno:2018gby}.
This question is directly relevant to neutron-star matter: sufficiently repulsive hyperon interactions can delay or suppress the appearance of hyperons and thereby alleviate the softening of the EOS associated with the hyperon puzzle.

Heavy-ion collisions provide an opportunity to put experimental constraints on strange-particle interactions at densities beyond those accessible in hypernuclei. 
A recent study~\cite{Ohnishi:2022nyw} parametrized a $\Lambda$ potential from $\chi$EFT calculations~\cite{Gerstung:2020ktv,Kohno:2018gby}, as a function of both density and momentum, in a microscopic transport model.
The resulting potential, which becomes strongly-repulsive at high densities, reproduces $\Lambda$ collective flow well. 
However, it is also noted that similarly good description can be obtained using a potential with weaker repulsion at high densities, indicating small sensitivity to the density-dependence of the potential, and that the flow is instead very sensitive to the momentum dependence of the strange interaction.
This mirrors the situation encountered for nucleons: extracting the density dependence of the interaction requires simultaneous control over its momentum dependence.

Beyond mean-field potentials, a challenge is also presented by elementary scattering cross sections and production rates involving strange hadrons.
Unlike nucleon--nucleon scattering, for which extensive experimental data exist, hyperon--nucleon and hyperon--hyperon interactions are only sparsely constrained.
A recent study in \texttt{UrQMD} provides a striking illustration of their importance.
After updating the $\Lambda N$ cross section using recent measurements from the CLAS collaboration~\cite{CLAS:2021gur}, Ref.~\cite{Reichert:2025rnw} found a large improvement in the rapidity dependence of the $\Lambda$ elliptic flow, bringing the calculation significantly closer to experimental data.

Heavy-ion collisions can therefore play a complementary role in constraining these poorly known interactions.
Differential measurements of strange-hadron yields and collective flow over collision energy, system size, rapidity, and transverse momentum probe different combinations of production cross sections, rescattering rates, and mean-field interactions.
Comparisons with transport calculations can provide indirect constraints on elementary inputs that are difficult to determine in dedicated scattering experiments.
At the same time, increasingly precise first-principles lattice-QCD calculations offer an independent route toward determining hyperon interactions.
The HAL QCD approach, for example, has extracted hyperon--nucleon and hyperon--hyperon potentials close to the physical point, from which scattering phase shifts and other low-energy observables can be calculated~\cite{HALQCD:2019wsz}.
Such calculations provide a particularly valuable complement to experiment in channels for which direct scattering data remain scarce.

Strangeness production below the elementary nucleon--nucleon threshold provides another sensitive probe of the dense stage of low-energy heavy-ion collisions.
Because the required energy cannot be supplied by a single $NN$ collision, subthreshold production proceeds through multi-step processes (\textit{e.g.}, $NN \to N \Lambda K^+$ can proceed through $NN\to N\Delta$ followed by $N\Delta \to N \Lambda K^+$) whose probability increases with density as more frequent scatterings in a dense environment increase the probability of accumulating the required energy. 
As a result, subthreshold production is sensitive to the compression achieved during the reaction, with larger yields correlated with higher densities, which in turn depend on the EOS.
This sensitivity has historically allowed subthreshold $K^+$ production to constrain the nuclear EOS~\cite{Fuchs:2003pc}.
Recent HADES measurements of subthreshold $\Lambda$ baryons~\cite{HADES:2022enx} demonstrate the feasibility of extending the EOS studies to subthreshold $\Lambda$ production, which nominally requires about $\sqrt{s_{\rm th}} \simeq 2.55~\rm GeV$ to be produced in a reaction such as $NN \to N\Lambda K$.
Particularly interesting in this context are new HADES measurements extending $\Lambda$ reconstruction to $\sqrt{s_{\rm NN}}=2.25~\mathrm{GeV}$, shown as preliminary during this conference. 
In this deep-subthreshold regime, strange-particle production necessarily proceeds through the collective reaction dynamics and multi-step processes, making it particularly sensitive to both the compression of the system and the microscopic production mechanisms implemented in transport calculations.

\subsection{Dynamics at low energies}

At $\sqrt{s_{\rm NN}}$ of several GeV, the reaction dynamics differs qualitatively from that encountered at higher energies.
The finite passage time of the two nuclei, substantial baryon stopping, and the continued presence of spectator matter throughout an appreciable fraction of the evolution all leave characteristic imprints on measured observables.

An important constraint on the longitudinal dynamics is provided by measurements of particle distributions.
For example, measurements of $dN/dy$ provide detailed information on baryon stopping, which not only directly influences the density and temperature histories relevant for EOS studies, but also depends on microscopic input to transport models that is often difficult to constrain, in particular hadronic cross sections in vacuum and their possible modifications in the medium.
The increasing availability of precise measurements -- \textit{e.g.}, proton $dN/dy$ at $\sqrt{s_{\rm NN}} = 3.2 ~ \rm GeV$~\cite{Labonte:2026oyq} or $\Lambda$ and hypernuclei yields over a series of fixed-target energies~\cite{Xie:2026dtx}, presented at this conference -- is therefore crucial for constraining these fundamental but still often poorly known inputs.

A particularly striking manifestation of the extended collision dynamics at these energies is triangular flow~$v_3$, measured by both HADES~\cite{HADES:2020lob} and STAR~\cite{STAR:2023duf}.
In contrast to ultrarelativistic energies, where $v_3$ is predominantly driven by event-by-event fluctuations of the initial geometry and is typically measured with respect to the third-order event plane~$\Psi_3$, at few-GeV energies a reaction-plane-correlated (\textit{i.e.}, $\Psi_1$-correlated) $v_3$ can arise from the time-dependent collision geometry and the interaction of the expanding matter with the spectator nucleons.
Transport calculations further indicate that this observable can exhibit significant sensitivity to the EOS~\cite{STAR:2023duf}.

Spectator effects can also qualitatively alter the interpretation of more familiar flow observables.
STAR measurements of number-of-constituent-quark~(NCQ) scaling of elliptic flow at $\sqrt{s_{\rm NN}}=3.0$--$4.5~\rm GeV$ show a pronounced breaking of the scaling at the lower energy and its gradual restoration with increasing $\sqrt{s_{\rm NN}}$~\cite{STAR:2025owm}.
A recent study~\cite{Reichert:2026dsb} demonstrated that an apparent separation between baryon and meson elliptic flow can arise naturally from spectator shadowing at low energies.
Thus, in the few-GeV regime, patterns in elliptic flow that at higher energies are often discussed in terms of the underlying microscopic degrees of freedom can also emerge from the space-time geometry of the collision.
This complicates the use of NCQ scaling, or its breaking, as a direct indicator of the onset or disappearance of partonic collectivity.

\subsection{Isospin dependence of nuclear interactions}

Any constraints on the high-density behavior of the EOS obtained from heavy-ion collisions should complement information from neutron stars and their mergers.
An important complication here is the poorly-constrained isospin-dependence of the EOS, which introduces substantial uncertainty when extrapolating from nearly symmetric heavy-ion matter to highly neutron-rich neutron-star matter~\cite{Yao:2023yda}.
To leading order in the isospin asymmetry, often quantified with $\delta \equiv (N_n - N_p)/(N_n + N_p)$, where $N_n$ and $N_p$ are neutron and proton number, respectively, the energy per baryon is~$E(n_B, \delta) \simeq E(n_B,0) + S(n_B)\delta^2 + \mathcal{O}(\delta^4)$. 
For collisions of stable heavy nuclei, such as gold, $\delta \simeq 0.2$, so that isovector effects enter as a correction to the dominant isoscalar dynamics and become difficult to isolate at several-GeV collision energies, where resonance production and rescattering further redistribute isospin.

The situation is more favorable at beam energies of several hundred MeV per nucleon, where observables such as neutron-to-proton flow ratios~\cite{Russotto:2011hq}, isospin diffusion~\cite{Zhang:2006vb,Zhang:2007hmv}, and charged-pion ratios retain greater sensitivity to the symmetry energy~\cite{SRIT:2021gcy,SpRIT:2020blg}.
Radioactive-beam facilities such as FRIB, RIKEN, and RAON further allow the isospin asymmetry of the colliding system to be varied systematically.
These measurements are especially valuable at densities of roughly $1.5$--$2.5n_0$, given that uncertainties in $\chi$EFT calculations grow rapidly above~$n_0$~\cite{Drischler:2021kxf} while neutron-star radii and tidal deformabilities provide limited sensitivity below~$2n_0$~\cite{Legred:2021hdx,Miller:2019nzo}.
They are further important for interpreting the high-density phase structure: features in the pressure or speed of sound inferred from nearly symmetric heavy-ion matter need not appear at the same density, or with the same strength, in neutron-rich matter.

\section{Summary}

Heavy-ion collisions provide the only experimental means of probing the QCD EOS over a broad range of baryon densities and temperatures. 
Realizing their full potential requires simultaneous progress in experiment, dynamical modeling, and the underlying physical inputs, which are essential not only for quantitative EOS constraints, but also for interpreting possible signatures of critical behavior and other structures in the QCD phase diagram.

At the same time, our expectations for the QCD phase diagram itself should not be regarded as settled. 
Recent work has questioned the conventional identification of the crossover near $T\simeq160~\rm MeV$ with the onset of a fully deconfined QGP, proposing instead an intermediate regime in which quark degrees of freedom are liberated while gluonic degrees of freedom remain confined, with full deconfinement occurring only at substantially higher temperature~\cite{Fujimoto:2025sxx}. 
To accommodate this and other exotic possibilities, experimental results should be interpreted without imposing \textit{a priori} assumptions about the structure of the QCD phase diagram.

\bibliographystyle{elsarticle-num}
\bibliography{sqm2026_ASorensen.bib}



\end{document}